\documentclass[aps,pra,epsfigure,onecolumn,superscriptaddress,notitlepage]{revtex4-1}
\usepackage[colorlinks=true,linkcolor=blue,urlcolor=blue,citecolor=blue,pdfusetitle]{hyperref}
\usepackage[utf8]{inputenc}
\usepackage[english]{babel}
\usepackage{amsmath}
\usepackage[caption = false]{subfig}
\usepackage{graphicx,epstopdf}
\usepackage{blindtext}
\usepackage{lipsum}
\usepackage{amsfonts}
\usepackage{bbm}
\usepackage{amssymb}
\usepackage{enumerate}
\usepackage{color}
\usepackage{latexsym}
\usepackage{physics}
\usepackage{times,txfonts}

\newcommand{\Ecal}{\mathcal{E}}
\newcommand{\Fcal}{\mathcal{F}}

\newcommand{\Scal}{\mathcal{S}}

\newcommand{\1}{\mathbbm{1}}

\newcommand{\interpro}[2]{\langle #1 | #2 \rangle}
\newcommand{\SubFig}[2]{\ref{#1}{\color{blue}#2}}

\definecolor{bluePoli}{cmyk}{0.4,0.1,0,0.4}
\definecolor{blueGreen}{RGB}{0, 179, 134}
\definecolor{brickred}{rgb}{0.8, 0.25, 0.33}
\definecolor{MyRed}{RGB}{255, 102, 102}

\definecolor{MyPurple}{RGB}{204, 0, 102}
\definecolor{MyRed}{RGB}{255, 102, 102}

\newcommand{\revisionRefA}[1]{{\color{black}#1}}
\newcommand{\revisionRefB}[1]{{\color{black}#1}}

\usepackage{ulem}

\usepackage{orcidlink}

\begin{document}
	
	\title{Encoding quantum bits in bound electronic states of a graphene nanotorus}
	
	\newcommand{\UFSCar}{Departamento de F\'{i}sica, Universidade Federal de S\~ao Carlos, Rodovia Washington Lu\'{i}s, km 235 - SP-310, 13565-905 S\~ao Carlos, SP, Brazil}
	\newcommand{\UFCA}{Centro de Ci\^{e}ncias e Tecnologia, Universidade Federal do Cariri,	63048-080, Juazeiro do Norte, Cear\'{a}, Brazil}
	\newcommand{\SU}{Department of Physics, Stockholm University, AlbaNova University Center 106 91 Stockholm, Sweden}
	\newcommand{\Nice}{Universit\'e C\^ote d'Azur, CNRS, Institut de Physique de Nice, 06560 Valbonne, France}
	\newcommand{\CSIC}{Instituto de Física Fundamental (IFF), Consejo Superior de Investigaciones Científicas (CSIC), Calle Serrano 113b, 28006 Madrid, Spain.}
	
	\author{J. Furtado~\orcidlink{0000-0002-1273-519X}}
	\email{job.furtado@ufca.edu.br}
	\affiliation{\UFCA}
	
	\author{A. C. A. Ramos}
	\email{antonio.ramos@ufca.edu.br}
	\affiliation{\UFCA}

	\author{J. E. G. Silva}
	\email{euclides.silva@ufca.edu.br}
	\affiliation{\UFCA}
	
	
	\author{R. Bachelard~\orcidlink{0000-0002-6026-509X}} 
	\email{romain@ufscar.br}
	\affiliation{\UFSCar}
	\affiliation{\Nice}
	
	\author{Alan C. Santos~\orcidlink{0000-0002-6989-7958}}
	\email{ac\_santos@iff.csic.es}
	\affiliation{\UFSCar}
	\affiliation{\SU}
	\affiliation{\CSIC}
	
	\date{\today}
	
	\begin{abstract}
		We propose to use the quantum states of an electron trapped on the inner surface of a graphene nanotorus to realize as a new kind of physical quantum bit, which can be used to encode quantum information. Fundamental tasks for quantum information processing, such as the qubit initialization and the implementation of arbitrary single qubit gates, can then be performed using external magnetic and electric fields. We also analyze the robustness of the device again systematic errors, which can be suppressed by a suitable choice of the external control fields. These findings open new  prospects for the development an alternative platform for quantum computing, the scalability of which remains to be determined.
	\end{abstract}
	
	\maketitle
	
	\section{Introduction}
	
	Quantum bits (qubits) constitute the most fundamental unit of information for quantum computation, exploiting the properties of two-level systems to create superposition of quantum states. Considering the potential advantage of quantum computers for solving hard problems through quantum algorithms, such as Shor's~\cite{Shor:94,Shor:97} and Simon's~\cite{Simon:94,Simon:97} ones, efforts have been dedicated to identifying suitable physical platforms to encode qubits. Up to date, qubits have been successfully encoded in degrees of freedom of light~\cite{Knill:01,Kok:07}, spins in nuclear magnetic resonance~\cite{Sarthour:Book}, spins in quantum dots~\cite{Loss:98}, semiconductor spins~\cite{Vandersypen:19}, and trapped ions~\cite{Cirac:95} among others~\cite{Nielsen:Book}. Recently, impressive progress has also been achieved with superconducting qubits~\cite{Arute:19,Wu-Pan:21}, driven by the interest of the community in proposing alternative kinds of physical qubits for quantum computation.
	
	On the other hand, the outstanding properties of two-dimensional materials sparked a revolution in physical science. In particular, the features of curved graphene have been exploited to devise new electronic devices, such as helical strips with chiral effects, also known as chiraltronics~\cite{DANDOLOFF2004233, PhysRevB.79.033404, PhysRevB.92.035440}, and Möbius strips with topological insulator properties~\cite{PhysRevB.80.195310}. Bridges connecting layers of graphene have also been proposed using a catenoid surface~\cite{DANDOLOFF20092667, PhysRevA.81.014102, SILVA2020126458}. These curved materials are described by an effective Dirac Hamiltonian for a massless and gapless electron in the single layer of graphene~\cite{katsnelson_2012}, and the surface curvature is associated with a geometric potential, the so-called da Costa potential~\cite{dacosta1, dacosta2}. Among the curved geometries of interest, the torus can be built by gluing the two edges of a nanotube and presents a rich phenomenology~\cite{AHLSKOG1999202, WANG200136, martel, Haddon97}. In particular, it exhibits a number of bound states, whose spectrum and geometry can be manipulated with external electric and magnetic fields~\cite{Silva:20}.
	
\begin{figure}[t!]
		\includegraphics[width=\linewidth]{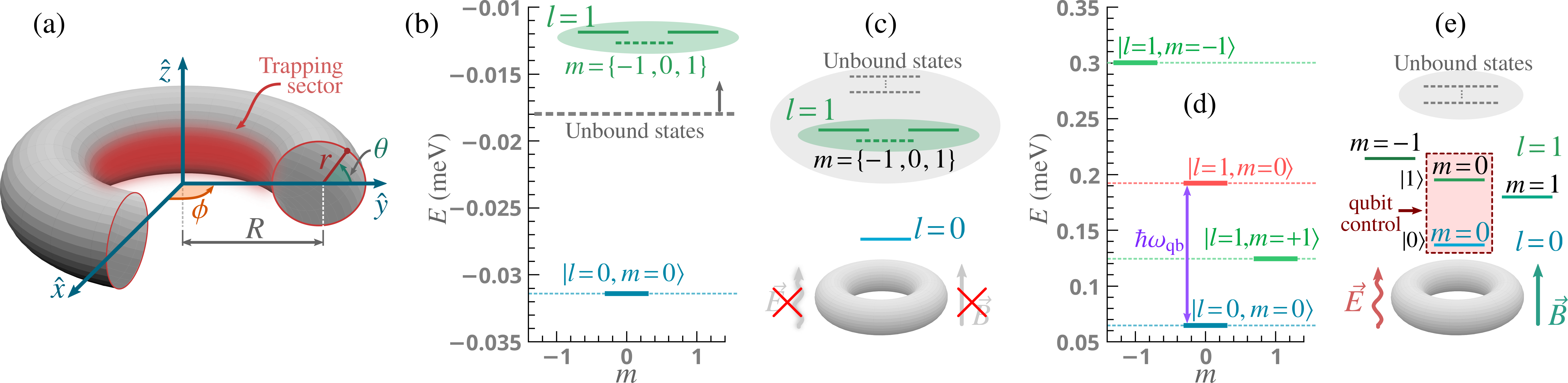}
		\caption{{(a) Schematic representation of the nanotorus submitted to a static magnetic field (used to prepare the qubit) and to a time-dependent electric field (used to control the qubit and implement gates). The trapping sector, where the electron is confined, is shown in red, for the case of the orbital angular momentum $m=0$ states. In Figs.~(b--e) we show the schematic representation of energy levels of interest to our work. (b,c) Energy diagram and bounded states of the system in absence of an external magnetic field, there exists a single bound state, corresponding to $\ket{l=0,m=0}$. (d) When applying an external magnetic field, geometrical effects induce a Zeeman-like breaking of the degeneracy of the states with $l=1$. (d) It is possible to see the emergence of a new bound state $\ket{l=1,m=0}$, used to realize our two-level system, arises. This allows one to encode a bit of quantum information in the energy subspace spanned by the states $\ket{l=\{0,1\},m=0}$. (e) In addition to the static magnetic field, a \textit{local} time-dependent (oscillating) electric field is used for qubit control by promoting transitions in the qubit subspace $\ket{l=\{0,1\},m=0}$, as we shall see soon.} }\label{Fig-Scheme}
	\end{figure}

	In this paper, we show how to encode a qubit in the quantum states of an electron confined on the surface of a graphene nanotorus, as illustrated in Fig.~\ref{Fig-Scheme}. The trapping potential induced by the curvature leads to a discrete set of quantum states, which are characterized by two quantum numbers, namely, the total angular momentum $l$ and its component along the quantization axis, $m$. We  show how the qubit can be initialized, and how gates can be implemented. Furthermore, the robustness of such information processing against systematic errors is characterized, showing how they can be mitigated by external control fields. Considering the number of techniques to produce and probe such a nanotorus (and nanorings, a similar geometry)~\cite{AHLSKOG1999202, WANG200136, martel, Haddon97,omachi2013initiation,Chen:05,Thorner:14,Kharissova:19,Lehr:14,omachi2011synthesis}, our proposal paves the way for the experimental realization of a new kind of qubit, under well-controlled conditions.
	
\section{Model and first quantization}
	
We here consider an electron confined on the surface of a graphene nanotorus, which is characterized by its minor radius $r$ and major radius $R$, see Fig.~\SubFig{Fig-Scheme}{a}. Thanks to the axial symmetry, the wavefunctions of the time-independent Schrödinger equation can be written as $\psi_{n,l,m}(\theta,\phi)=\chi_{n,l}(\theta)e^{im\phi}$, where $\phi$ is the angle around the torus axis of revolution, and $\theta$ the one around the tube. As we shall see, $m$ is the quantum number associated with the component of the angular momentum in direction $\hat{Z}$, while the angular component of the wavefunction $\chi_{n,l}(\theta)$ accounts for the quantum numbers related to the electronic energy level $n$ and the total angular momentum of the system $l$. The latter component satisfies the following time-independent Schr\"{o}dinger equation~\cite{Silva:20}:
	\begin{align}\label{e12}
	-\frac{\hbar^2}{2m^{*}r^2}\frac{d^2\chi(\theta)}{d\theta^2} +V_{\text{bare}}(\theta)\chi(\theta) =\epsilon\chi(\theta), 
	\end{align}
	where the potential $V_{\text{bare}}$ reads
	\begin{align}
	V_{\text{bare}}(\theta)&=\frac{\hbar^2}{2m^{*}r^2(R+r\cos\theta)^2}\left[-\frac{R^2}{4}+r^2m^2+\frac{r^2\sin^2\theta}{4} + \frac{r(R\cos\theta+r)}{2}\right] , \label{Eq-Vbare}
	\end{align}
	with $m^{*}=0.3m_0$ the effective mass of the electron on the torus surface, and $m_0$ the rest mass of the electron~\cite{Haddon97}. Under a suitable choice of the parameters $R$ and $r$, a single confined state exists, for which the probability density function reveals a ``trapping sector" in the inner region of the nanotorus ($\theta=\pi$). As can be observed in Fig.~\SubFig{Fig-Scheme}{a}, its width at half height is $\delta \theta\approx \pi$, so the electronic wavefunction forms a ring located around $\theta=\pi$~\cite{Silva:20}. As for the global multiplicative term in Eq.~\eqref{Eq-Vbare}, which depends on the effective mass of the electron and on the torus parameters, it sets the energy scale of our system. The first term in the brackets stands for Da Costa potential, which results from the squeezing of the electron wavefunction on the curved surface~\cite{dacosta1, dacosta2}, and its contribution is to locate the electron in the inner region of the nanotorus. The other three remaining contributions are centrifugal ones, and they result from the geometry of the system,  see~\cite{Silva:20} for more details.
	
	However, in presence of external fields, whether electric or magnetic, the potential $V_{\text{bare}}$ must be substituted by the effective potential $V(\theta)=V_{\text{bare}}(\theta)+V_{E}(\theta)+V_{B}(\theta)$, in which $V_{E}(\theta)$ and $V_{B}(\theta)$ account for electric and magnetic field contributions, respectively:
	\begin{align}
	V_{E}(\theta)&=-eEr\sin\theta, ~~
	V_{B}(\theta)=\frac{e^2B^2}{8m^{*}}(R+r\cos\theta)^2-\frac{me\hbar B}{2m^{*}}. \label{Eq-VMagnetic}
	\end{align}
	The Hamiltonian of the system in presence of these fields then reads
	\begin{eqnarray}
	H(\theta)=-\frac{\hbar^2}{2m^{*}r^2}\frac{d^2}{d\theta^2} + V(\theta). \label{Eq-Hamiltonian}
	\end{eqnarray}
	
	These external fields can drastically alter the confinement potential of the electron on the nanotorus surface, leading to new confined electronic near-ground states. In what follows, we focus on the set of states associated with the total angular momentum $l = \{0,1\}$, since they are already suitable candidates to encode a quantum bit of information.

	\subsection{Electronic near-ground states}

    \revisionRefA{First of all, let us motivate that the choice for values of $r=350 \AA$ and $R=900 \AA$ or $R=3600 \AA$ are in agreement with experimental data (first reported in the literature in \cite{AHLSKOG1999202}), and it is also important to highlight here that the value for the minor radius $r$ can be related to the radius of multi-walled carbon nanotubes (MWCNT) (see Ref.~\cite{Saira:15_review} for further details and discussions). Also, since we are considering MWCNT we have used the Schrödinger equation to describe the electron's dynamics, due to the quadratic dispersion relation associated with the multi-layer structure, as presented previously. In the case of a monolayer torus, the massless Dirac equation should be considered instead.}
	
	\revisionRefA{In Fig.~\SubFig{Fig-Scheme}{b--d}}, we present the changes in the energy level scheme induced by the external magnetic field (from $B=0$T to $B=0.45$T), with the lowest-lying states associated with $l = \{0,1\}$ represented. {As a first relevant remark about our system, we take into account the fact that our potential is not an infinite potential well, what means that the solutions of the time-independent Schrödinger equation can be classified as \textit{bounded} and \textit{unbounded} states. It means that solutions of Eq.~\eqref{e12} with energy higher than a given threshold $E_{\mathrm{trap}}$ (trapping energy) are unbounded state and we do not use them for store quantum information. In fact, as shown in Figs.~\SubFig{Fig-Scheme}{b,c}, in absence of external magnetic field we observe the existence of a single bound state corresponding to $\ket{l=0,m=0}$, as the trapping energy for the parameters considered is $E_{\mathrm{trap}}(B=0) \approx - 0,0283~meV$, and all states with $l\leq 1$ are solutions of the Eq.~\eqref{e12} with energy higher than $E_{\mathrm{trap}}(B=0)$. Therefore, to encode quantum information we need to promote the ``trapping" of new states. It is in fact possible due to the positive contribution of the magnetic field to the potential as given in Eq.~\eqref{Eq-VMagnetic}, which allows us to increase the energy $E_{\mathrm{trap}}$. As depicted in Figs.~\SubFig{Fig-Scheme}{d,e}, by applying an external magnetic field strong enough the potential is drastically modify and the new trapping energy becomes $E_{\mathrm{trap}}(B=0.45T) \approx 0.3019~meV$, what allows us to confine the new set of states given by $\ket{l=1,m=0,\pm 1}$. In addition to the ``trapping" new bound states, we also highlight the positive effect of the static magnetic field related to a Zeeman-like effect in the subspace $\{\ket{l=1,m=\pm 1}\}$. Mainly due to the last term in Eq.~\eqref{Eq-VMagnetic}, the magnetic field breaks the degeneracy of such a subspace, as illustrated in Fig.~\SubFig{Fig-Scheme}{e}.}
	
	We point out that the effect of the external electric field is to increase the number of confined states also increases and becomes spatially displaced from the nanotorus inner radius. Nevertheless, these states are doubly degenerate to $m= \{\pm 1, \pm 2, \cdots\}$, a situation which is undesirable for the initialization of the qubit state. Indeed, degenerate states are not adequate for quantum information encoding, since multiple states may become populated. This motivates the application of a static magnetic field on the nanotorus to achieve a proper qubit. We emphasize that the breaking of degeneracy by the magnetic field is not exactly a Zeeman effect, in the sense that the field drastically modifies the geometric potential, in addition to shifting the energy of the associated levels.

\section{Electronic qubit encoding}
	
Let us adopt DiVincenzo's approach to characterize our system as a qubit suitable to encode quantum information~\cite{Divincenzo:00}. We here focus on the two following criteria: \textit{A physical system (1) with well characterized qubit (i.e., two-level system), and (2) which can be coherently controlled, is a candidate for quantum information processing}. The additional DiVicenzo's criteria concern the scalability, decoherence effects and the readability of the quantum information encoded in the qubits, which we discuss briefly in the conclusion. 
	
\begin{figure}
	\includegraphics[width=\linewidth]{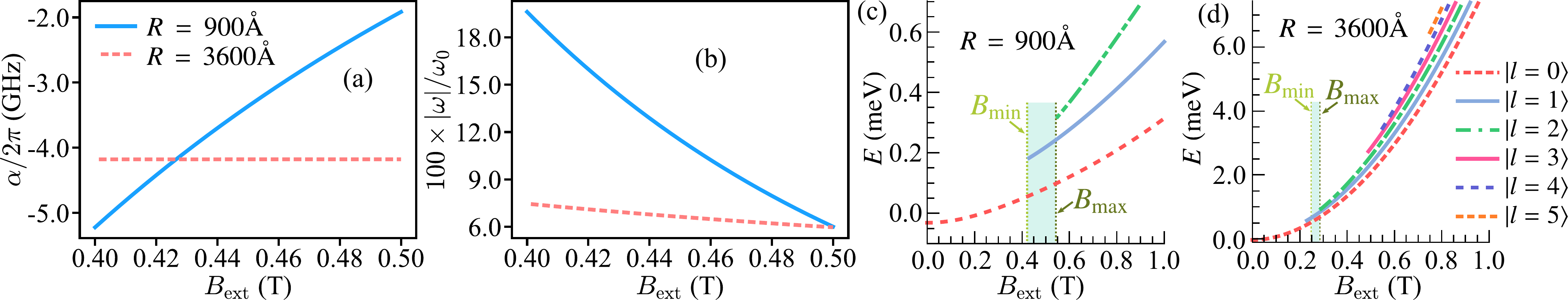}
	\caption{(a) Anharmonicity parameter $\alpha$ and (b) the ratio $|\alpha|/\omega_{0}$ for two different values of the external radius $R$, for $r=350\AA$. (c,d) Energy levels (with $m=0$) of an electron confined on the inner radius of a nanotorus as function of the external magnetic field $B$ (in Tesla) for two configurations: (c) $r=350\AA$ and $R=900\AA$, and for (d) $r=350\AA$ and $R=3600\AA$. In both cases, the first and second curves, from bottom to top, correspond to the fundamental level $\ket{l=0,m=0}$ and first excited one $\ket{l=1,m=0}$, respectively. The other curves correspond to other energy levels, which are undesirable in the present scheme to encode a qubit.}\label{Fig-Anharmonicity}
\end{figure}

	We first consider the case when an external magnetic field is applied over the system and the orbital angular momentum $m$ is taken to be zero. The $m\!=\!0$ states are often considered, for example, for quantum computing with trapped ions~\cite{Leibfried:03,barreiro2011open,Hu:18,Hu:20-a}. Returning to Eq.~\eqref{Eq-Hamiltonian} and doing an expansion around the trapping region ($\theta\!\approx\!\pi$), one obtains 
	\begin{align}
	V_{\text{bare}}(\theta_{\pi})+V_{B}(\theta_{\pi}) = \frac{\beta^2}{2m^{*}} \theta_{\pi}^2 + \delta \theta_{\pi}^4 + \epsilon,
	\end{align}
	where we have introduced the angular variable $\theta_{\pi}\!=\!\theta-\pi$. The coefficients $\beta$, $\delta$ and $\epsilon$ are given by
	\begin{subequations}
		\begin{align}
		\delta&=-\frac{1}{96m^{*}}\left[\frac{\hbar^2r(R+8r)}{(R-r)^4}+e^2B^2r(R-4r)\right],\\
		\beta^2&=\frac{r}{4}\left[\frac{\hbar^2}{(R-r)^3}+e^2B^2(R-r)\right],\\
		\epsilon&=\frac{1}{8m^{*}(R-r)^2}\left[e^2B^2(R-r)^4-\frac{\hbar^2\left(R^2-2Rr+2r^2\right)}{r^2}\right].
		\end{align}
		\label{Eq-Parameters}
	\end{subequations}
	The first quantization is realized by identifying the canonical variables of position, $\theta_{\pi}\rightarrow\hat{\theta}_{\pi}$, and momentum $\hat{p}_{\theta}\!=\!(i/r)d(\bullet)/d\theta_{\pi}$, as operators. In this first quantization picture, the Hamiltonian of the system reads
	\begin{align}
	\hat{H}_{0} = -\frac{\hbar^2}{2m^{*}r^2}\frac{d^2}{d\theta_{\pi}^2} + \epsilon + \frac{\beta^2}{2m^{*}} \hat{\theta}_{\pi}^2 + \delta \hat{\theta}_{\pi}^4 . \label{H0}
	\end{align} 
	This Hamiltonian describes a modified harmonic potential with anharmonicity $\delta$, which is reminiscent of the one obtained for superconducting qubits~\cite{Blais:04,Koch:07,Krantz:19,Rasmussen:21}. In particular, the change of variables $\hat{\theta}_{\pi} \rightarrow \hat{q}/r$ leads to
	\begin{align}
	\hat{H}_{0} = \frac{\hat{p}^2}{2m^{\ast}} + \frac{m^{\ast}\omega_{0}^2}{2}\hat{q}^2 + \frac{\delta}{r^4} \hat{q}^4 + \epsilon \1, \label{H0-positionMomentum}
	\end{align} 
	with $\omega_{0}=|\beta|/m^{*}r$. The connection with artificial superconducting atoms is made more straightforward by defining the creation and annihilation operators, $\hat{a}^{\dagger} = \left( \hat{q} - id/d\hat{q} \right)/\sqrt{2}$ and $\hat{a} = \left( \hat{q} + id/d\hat{q} \right)/\sqrt{2}$ respectively, so we obtain the anharmonic oscillator Hamiltonian:
	\begin{align}
	\hat{H}_{0} = \1 E_{0} +\hbar \omega_{0} \hat{a}^{\dagger}\hat{a}  + \alpha \left(\hat{a} + \hat{a}^{\dagger}\right)^4 , 
	\end{align} 
	with $E_{0}\!=\!\epsilon + \hbar \omega_{0}/2$ and $\alpha = ( \hbar / 2m^{\ast}\omega_{0} r^2 )^2 \delta$. The first two terms correspond to the quantum harmonic oscillator with ground-state energy $E_0$ and a spectrum of energies separated by quanta of energy $\hbar \omega_{0}$. The third term of the above equation stands for the anharmonicity parameter. 
	For the regime of parameters used in Fig.~\ref{Fig-Scheme}, the harmonic contribution to energy is much bigger than the anharmonic one ($\omega_0 \gg |\alpha|$), as shown in \revisionRefA{Figs.~\SubFig{Fig-Anharmonicity}{a,b}}. Then under the rotating wave approximation we write $(\hat{a} + \hat{a}^{\dagger})^4\approx 3\1 + 12\hat{a}^{\dagger}\hat{a} + 6 \hat{a}^{\dagger}\hat{a}^{\dagger}\hat{a}\hat{a}$. Thus, we can rewrite the above Hamiltonian in a more convenient way as
	\begin{align}
		\hat{H}_{0} \approx \hbar \omega \hat{a}^{\dagger}\hat{a}  + 6\alpha\hat{a}^{\dagger}\hat{a}^{\dagger}\hat{a}\hat{a} , \label{H_0}
	\end{align}
	where $\omega = \omega_0 + 12\alpha$. We hereafter refer to this Hamiltonian as our \textit{bare Hamiltonian}, which defines the typical properties of our system, such as the qubit energy scale and qubit transition frequency. \revisionRefA{We can use this result to estimate the qubit frequency. For instance, consider the parameters given by Fig.~\SubFig{Fig-Anharmonicity}{c}, where the nanotorus has geometry parameters $r=350\AA$ and $R=900\AA$ and a local local magnetic field of $0.45~T$, we can estimate $\omega/2\pi \approx 25.9$~GHz.}
	
	As an alternative to genuine two-level systems, such as spin and photon polarization qubits, one can exploit the large anharmonicity to induce energy shifts in the upper-excited levels, so the system operates effectively as a two-level system. Indeed, when driving the lowest excited state, the probability to gain a second quantum of energy is negligible. This scheme is often used in artificial atoms, such as in superconducting qubits~\cite{Krantz:19,Rasmussen:21}, systems where the self-Hamiltonian is equivalent to the one of Eq.~\eqref{H_0}. In the case of the nanotorus, increasing the external magnetic field $B$ allows one to produce an effective two-level system that spans the subspace $\Scal_{\mathrm{qb}}=\{\ket{l=0,m=0},\ket{l=1,m=0}\}$, with the main quantum number $m\!=\!0$ and $l=\{0,1\}$, see Fig.~\ref{Fig-Scheme}. Now assuming that the system only explores the two energy levels of $\Scal_{\mathrm{qb}}$, the bosonic operators $\hat a$ and $\hat a^\dagger$ can be substituted by $\hat{\sigma}_{-}$ and $\hat{\sigma}_{+}$, respectively, the Pauli matrices defined as $\hat{\sigma}_{-}=\ket{l=0,m=0}\bra{l=1,m=0}$ and $\hat{\sigma}_{+}=(\hat{\sigma}_{-})^{\dagger}$. The Hamiltonian for the qubit then reads 
	\begin{align}
	\hat{H}_{\mathrm{qb}} = \hbar \omega \hat{\sigma}_{+}\hat{\sigma}_{-} + \alpha\left(\hat{\sigma}_{+} + \hat{\sigma}_{-}\right)^4 + \1 E_{0} .
	\end{align} 
	
	Moreover, since the second term in the above equation can be rewritten as $\alpha\left(\hat{\sigma}_{+} + \hat{\sigma}_{-}\right)^4=\alpha\1$, we simplify the above Hamiltonian by redefining the zeroth energy level to get 
	\begin{align}
	\hat{H}_{\mathrm{qb}} = \hbar \omega \hat{\sigma}_{+}\hat{\sigma}_{-}  . 
	\end{align}
	We stress that this Hamiltonian is valid only when transitions occur only inside the subspace $\Scal_{\mathrm{qb}}$, where no change in the main quantum number is promoted. With a stronger magnetic field, new bound states with $m=0$ will appear, so the system will turn into an effective multilevel atom. 
	
	\revisionRefA{It is worth to mention that the other states with $m = \pm 1$, in principle, also can be used as useful states to encode qubits. In fact, thanks to the break of degeneracy aforementioned, an external drive able to promote transitions $\ket{l=0,m=0} \rightarrow \ket{l=1,m=\pm1}$ can also lead to a dynamics for different subspaces composed of two independent states. For example, if we drive at transition $\ket{l=0,m=0} \rightarrow \ket{l=1,m=+1}$, we do not populate the states $\ket{l=1,m=-1}$ and $\ket{l=1,m=0}$ because they are off-resonant with such a transition. Therefore, we create a qubit with the subspace $\{\ket{0} =\ket{l=0,m=0} ,\ket{1} = \ket{l=1,m=+1}\}$.}
	
	\section{Qubit initialization and control}
	
	As mentioned before, when the magnetic field is below a certain value $B_{\mathrm{ini}}$, hereafter called initialization static magnetic field, the system presents a single low-lying state for which $m=0$. The others states have a number $m\neq 0$, as illustrated in Fig.~\ref{Fig-Scheme}. In this case, an electron trapped in the nanotorus tends to populate this specific state, since it has the lowest energy. The qubit initialization is then achieved by introducing an external magnetic field which is strong enough to turn the electronic state $\ket{l=1,m=0}$ into a bound state, yet not so strong that the subspace corresponding to $m=0$ would contain more than two states. Hence, the system behaves as an effective two-level system. Since the system has previously been initialized in the state $\ket{l=0,m=0}$, and since the magnetic field variation does not promote any transition, the system remains in the state $\ket{\psi(0)} = \ket{l=0,m=0}$, which concludes the initialization of the qubit.

 \revisionRefA{In Figs.~\SubFig{Fig-Anharmonicity}{b,c}}, we present the energy of the bound states as a function of the magnetic field amplitude. The values of $B_{\mathrm{min}}$ and $B_{\mathrm{max}}$ are illustrated by vertical dotted lines, for two different values of the external radius $R$. Indeed, the precise range of values for the magnetic field required to initialize the qubit depends on the specific geometry of the nanotorus, through the minor $r$ and major radius $R$.
	
 \subsection{Qubit control and universal single-qubit quantum gates}
	
As for the majority of quantum emitters, the driving of a specific transition can be achieved using an external electric field, using a local field to control the qubit. The extra electric field $\vec{E}=E\hat{Z}$ results in an additional contribution to Hamiltonian~\eqref{Eq-Hamiltonian}, which reads
	\begin{align}
	\hat{H}_{E} = erE \sqrt{\frac{\hbar}{2r|\beta|}} \left[\left(\hat{a} + \hat{a}^{\dagger}\right) -\frac{\hbar}{12r|\beta|} \left(\hat{a} + \hat{a}^{\dagger}\right)^3\right].
	\label{Eq-H_extEMultilevel}
	\end{align}
	Considering again an external magnetic field such that we have only the low-lying states $\ket{n}\!=\!\ket{l=n,m=0}$, we can replace the operators $\hat{a}$ and $\hat{a}^{\dagger}$ by $\hat{\sigma}_{-}$ and $\hat{\sigma}_{+}$ in Eq.~\eqref{Eq-H_extEMultilevel}, which leads to
	\begin{align}
	\hat{H}_{E} = \hbar \Omega(E) (\hat{\sigma}_{+} + \hat{\sigma}_{-}).
	\end{align}
	We have used that $\left(\hat{\sigma}_{-} + \hat{\sigma}_{+}\right)^3=\left(\hat{\sigma}_{-} + \hat{\sigma}_{+}\right)$ and introduced the frequency $\Omega(E) = \mu E/\hbar$, with $\mu$ being given by
	\begin{align}
	\mu = er \left[\sqrt{\frac{\hbar}{2r|\beta|}} -\frac{1}{6}\sqrt{\left(\frac{\hbar}{2r|\beta|}\right)^3} \right] . \label{Eq-Mu}
	\end{align}
	
	Since the quantity $\hbar/r|\beta|$ is dimensionless, $\mu$ has dimension of a `charge-distance', i.e., the dimension of an electric dipole moment (see Appendix~\ref{ApSec-DimenAnaly} for further details). Furthermore, the amplitude of $\Omega(E)$ here depends on both the electric and magnetic fields, in addition to other parameters related to the system geometry. The controllability of the system is thus strongly related to the control of the driving electric field $E$, once the other parameters have been fixed. In particular, by considering a time-dependent oscillating \revisionRefB{driving} field $E$ given by
	\begin{align}
	E = E_{0} \cos(\revisionRefB{\omega_{\mathrm{d}}}t+\phi),
	\end{align}
	with $\phi$ an adjustable phase, the driving Hamiltonian becomes
	\begin{align}
	H_{\mathrm{dv}}(t) = \hbar \omega \sigma^{+}\sigma^{-} + \hbar \Omega(E_0) \cos(\revisionRefB{\omega_{\mathrm{d}}}t+\phi) (\sigma^{+}+\sigma^{-}).
	\end{align}
	
	Finally, let us derive the system effective dynamics close to resonance. According to Schrödinger equation, the system evolves as
	\begin{align}
	i\hbar \ket*{\dot{\psi}(t)} = H_{\mathrm{dv}}(t)\ket{\psi(t)},
	\end{align}
	since no external dissipative channel is taken into account. In order to explore the two-level nature of the system, we then consider the case of a resonant driving ($\Omega(E_0) \ll \omega$). 
	Furthermore, we move to the rotating frame of the external field and use the Rotating Wave Approximation, so the dynamics of the system is dictated by the driving effective Hamiltonian (see Appendix~\ref{ApSec-EffecDyn})
	\begin{align}
	H_{\text{eff}} = \hbar \Delta \sigma_{+}\sigma_{-} + \frac{\hbar \Omega(E_0)}{2} \left( e^{i\phi} \sigma_{-} + e^{-i\phi} \sigma_{+}\right), \label{Ap-Heff}
	\end{align}
	where $\Delta\!=\! \omega - \revisionRefB{\omega_{\mathrm{d}}}$ is the detuning between the qubit transition frequency and the external field frequency. The above Hamiltonian, where the phase $\phi$ is controlled through the driving field, can then be used to implement \textit{arbitrary} single qubit gates~\cite{Santos:18-a}, so one can conclude that the di Vincenzo's criteria to use a physical qubit for quantum computation are fulfilled.	
	
	\subsection{Preparation of an arbitrary input state}
	
	In this section we show how to choose adequately the parameters of the effective Hamiltonian $H_{\text{eff}}$ in order to prepare an arbitrary input state given by
	\begin{align}
	\ket{\psi(\theta,\eta)} = \sin\left( \theta/2 \right)\ket{0} + e^{i\eta} \cos\left( \theta/2 \right)\ket{1}, \label{Eq-psi-gen}
	\end{align}
	for any $\theta\in(0,\pi]$ and $\eta \in [0,\pi]$. The system is initially prepared in the state $\ket{0}$ using the procedure discussed previously. Then, using a resonant \revisionRefB{driving} field $(\revisionRefB{\omega_{\mathrm{d}}}=\omega)$, the system evolves into $\ket{\Psi(t,\phi)}= e^{-\frac{i}{\hbar}H_{\text{eff}} t}\ket{0}$, which reads
	\begin{align}
	\ket{\Psi(t,\phi)} &= \cos\left( \frac{t\Omega}{2} \right)\ket{0} + e^{-i \left(\phi + \frac{\pi}{2}\right)}\sin\left( \frac{t\Omega}{2} \right)\ket{1} .
	\end{align}
	
\begin{figure}[t!]
	\includegraphics[width=0.75\linewidth]{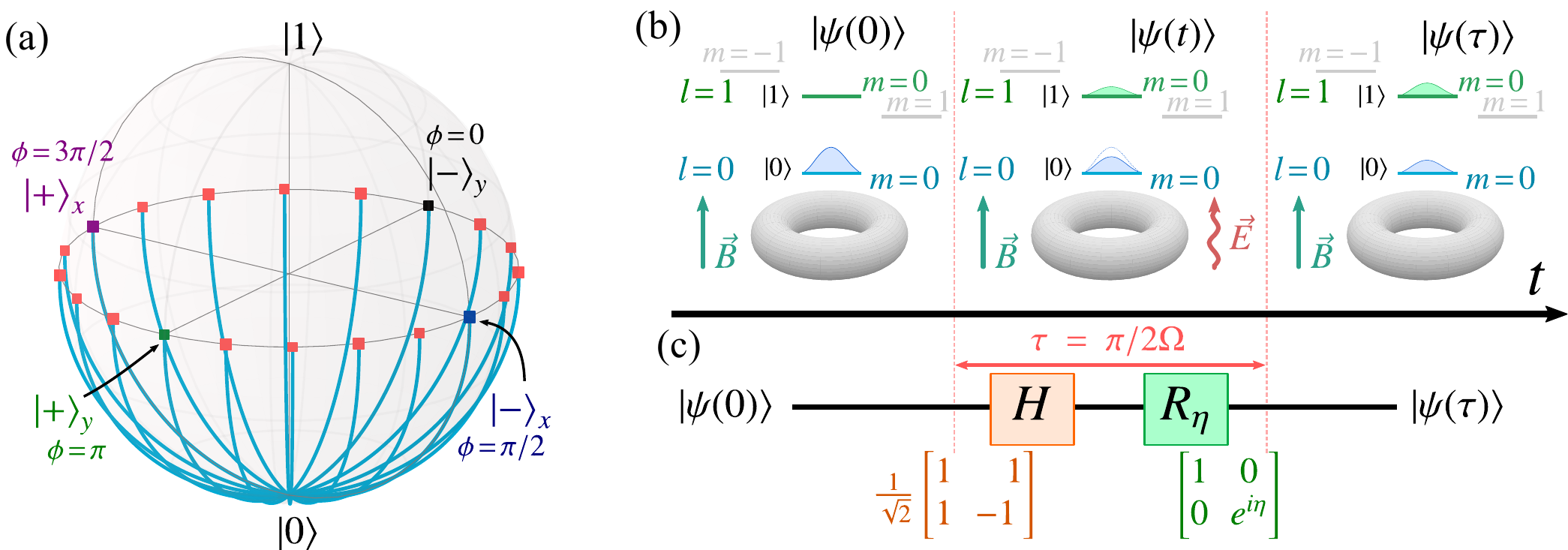}
	\caption{(a) Trajectories on the Bloch sphere obtained for different choices of the parameter $\phi$. The total evolution time $\tau = \pi/2\Omega$ brings the system to one of the state in the equator of the sphere, where any state in $\Ecal$ can be prepared. The states in the $x$- and $y$-axis of the Bloch sphere are defined as $\ket{\pm}_x=(\ket{0}\pm\ket{1})/\sqrt{2}$ and $\ket{\pm}_y=(\ket{0}\pm i \ket{1})/\sqrt{2}$. (b) Schematic representation of each step in the implementation of the quantum circuit shown in (c), where a magnetic field is used to initialize the system in the ground state $\ket{0}$, followed by the coherent control using the external electric field which drives the system to a desired final state. In (c) we show the quantum gates Hadamard ($H$) and phase gate ($R_{\eta}$), satisfying $H\ket{n}=(\ket{0}+(-1)^{n}\ket{1})/\sqrt{2}$, with $n\in\{0,1\}$, and $R_{\eta}(a\ket{0} + b\ket{1})=a\ket{0} + be^{i\eta}\ket{1}$.}
	\label{Fig-BlochSphere}
\end{figure}
	
	By comparing the above equation with Eq.~\eqref{Eq-psi-gen}, it becomes clear the total evolution time and the phase $\phi$ can be chosen to achieve any state of the form \eqref{Eq-psi-gen}. As an illustration, we show in Fig.~\SubFig{Fig-BlochSphere}{a} the trajectory of the quantum state on the Bloch sphere, driven by Hamiltonian~\eqref{Ap-Heff}. After initializing the system in the state $\ket{0}$, we set the total evolution time as $\tau = \pi/2\Omega$ such that the final state is one of the set of quantum states $\Ecal$ at the equator of the Bloch sphere, which is a set of states with maximum quantum coherence in basis $\{\ket{0},\ket{1}\}$~\cite{Streltsov:17}. That is, the set $\Ecal$ of all states of the form $\ket{\xi(\eta)}\!=\!(\ket{0} + e^{i\eta}\ket{1})/\sqrt{2}$. Hence, by tuning the parameter $\phi$, we can prepare any state in $\Ecal$. In Fig.~\SubFig{Fig-BlochSphere}{b}, we show sequential steps of the implementation for each evolution considered in Fig.~\SubFig{Fig-BlochSphere}{a}. In Fig.~\SubFig{Fig-BlochSphere}{c} we present the equivalent quantum circuit, which corresponds to the Hadamard and phase gates, two genuine quantum gates which do not have an equivalent in classical computation~\cite{Nielsen:Book}.
	
 \section{Systematic errors and gate fidelity}
	
For a fixed set of values of the electric $E_{0}$ and magnetic fields $B_{0}$, and total evolution time $\Delta t$, the evolution operator $U(t)$ of the system driven by Hamiltonian~\eqref{Ap-Heff} characterizes a unitary quantum gate $G$~\cite{Nielsen:Book,Santos:18-a}. Therefore, uncontrollable or unpredicted fluctuations in the electric and magnetic field around their reference values, $E_{0}$ and $B_{0}$, may directly affect the qubit control in several ways. In fact, in our system, according to Eqs.~\eqref{Eq-Parameters},~\eqref{Eq-Mu} and~\eqref{Ap-Heff}, the fluctuations of the magnetic field have a direct impact on the qubit energy (and thus on its frequency), as well as on the effective electric dipole moment. Differently, errors on the electric field amplitude only affects the control of the qubit. In this section we discuss the performance of the gate implementation in presence of such systematic errors.
	
\begin{figure*}[t!]
\includegraphics[width=\linewidth]{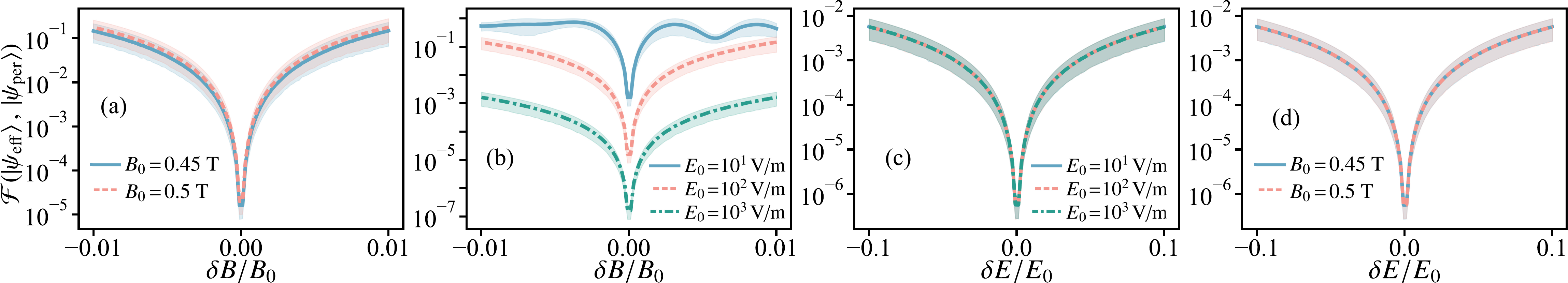}
\caption{(a--d) Average infidelity of the Hadamard gate under systematic errors in the magnetic and electric fields. (a) Average infidelity as a function of the error in the magnetic field, computed from $N=10,000$ random input states, for two different values of $B_0$ and for $E_0=100$~V/m. (b) Infidelity as a function of the error in the magnetic field, for different electric field amplitudes and for $B_0=0.45$~T: Increasing the electric field allows on to mitigate the gate error. (c) Average infidelity as a function of the error in the electric field, for different values of $E_0$ and for $B_0=0.45$~T. (d) Same as (c), but for different values of the magnetic field and for $E_0=1,000$~V/m: No reduction of the infidelity is observed within the range defined in Fig.~\SubFig{Fig-Anharmonicity}{c}. In all graphs, the parameters considered in Fig.~\SubFig{Fig-Anharmonicity}{c} were used, and the shadowed areas represent the fluctuations of the fidelity obtained from the maximum deviation $\delta \Fcal = \max_{\{\psi_\mathrm{inp}\}} \Fcal(\ket{\psi_\mathrm{eff}}, \ket*{\psi_\mathrm{per}})$ over the random input states.}
		\label{Fig-Systematic}
	\end{figure*}
	
We here consider an analysis of the systematic error based on the deviation between a desired quantum gate $G$ (target unitary operation) and the real operation implemented through the perturbed Hamiltonian. We call the unperturbed and perturbed Hamiltonians $H_{\mathrm{eff}}$ and $H_{\mathrm{per}}$, respectively, where $H_{\mathrm{per}}$ is obtained from Eq.~\eqref{Ap-Heff} by assuming deviations $\delta B$ and $\delta E$ in the magnetic and electric fields, respectively. We then let the system evolve under the action of $H_{\mathrm{per}}$ and compare the outcome state $\ket{\psi_\mathrm{eff}}$ with the state $\ket*{\psi_\mathrm{per}}$ expected from the ideal dynamics of $H_{\text{eff}}$, during a time $\Delta t$ necessary to implement the gate. In this scenario, each dynamics is dictated by its corresponding evolution operator
	\begin{align}
		U_{\mathrm{eff}/\mathrm{per}}(\Delta t) = \exp\left( \frac{1}{i \hbar} H_{\mathrm{eff}/\mathrm{per}} \Delta t \right),
	\end{align}
	where the unperturbed dynamics gives $U_{\mathrm{eff}}(\Delta t)=G$, while the perturbed evolution usually leads to $U_{\mathrm{per}}(\Delta t) \neq G$. To quantify the robustness of the system against these systematic errors, we compute the \textit{infidelity} as given by the Bures metric for pure states~\cite{Nielsen:06}
	\begin{align}
		\Fcal(\ket{\psi_\mathrm{eff}}, \ket*{\psi_\mathrm{per}}) = 1 - |\interpro{\psi_\mathrm{eff}}{\psi_\mathrm{per}}|^2 ,
	\end{align}
	with $\ket*{\psi_{\mathrm{eff}/\mathrm{per}}}=U_{\mathrm{eff}/\mathrm{per}}(\Delta t)\ket*{\psi_\mathrm{input}}$ and $\ket*{\psi_\mathrm{input}}$ an arbitrary input state. Therefore, $\Fcal(\ket{\psi_\mathrm{eff}}, \ket*{\psi_\mathrm{per}})$ ranges from $0$ (no error) to $1$ (maximum error). Since we need to implement single qubit gates on arbitrary input states, the error analysis is realized using the average over many input states of a given operation. We implement each dynamics and take the average of $\Fcal(\ket{\psi_\mathrm{eff}}, \ket*{\psi_\mathrm{per}})$ over $N$ random input states $\ket*{\psi_\mathrm{input}}$. 
	
Let us here consider the implementation of a Hadamard gate (a genuinely quantum gate), and compute the average infidelity as a function of the fluctuation amplitude of the magnetic field, $\delta B/B_0$, where $B_0$ is a reference value. A Hadamard gate is implemented in a time interval $\Delta t$ by choosing $\Delta = \Omega(E_0)/2 = \pi / (2\sqrt{2}\Delta t)$ in Eq.~\eqref{Ap-Heff}~\cite{Santos:18-a}. Therefore, we set the \revisionRefB{driving} field frequency $\revisionRefB{\omega_{\mathrm{d}}}$ and the reference magnetic and electric field $B_0$ and $E_0$ accordingly.
	
In Fig.~\SubFig{Fig-Systematic}{a} we show the behaviour of the infidelity as a function of $\delta B/B_0$ and for two different values of magnetic field $B_0$. This systematic error induces a change in the natural qubit frequency $\omega$, so the detuning $\Delta$ is in turn affected (since $\revisionRefB{\omega_{\mathrm{d}}}$ is kept fixed); then, the condition $\Delta = \pi / (2\sqrt{2}\Delta t)$ is no longer satisfied. In the cases considered here, small deviations in $B_0$ (less than $1\%$) are enough to induce a significant infidelity in the gate implementation. Note that errors of this magnitude are within reach of several experiments~\cite{fei1997precision, goerbig2011electronic}, especially in state-of-the-art setups~\cite{Gross:16,Miller:17,Thiele:18} where the control over the magnetic field is of order of a few $\mu$T. However, as shown in Fig.~\SubFig{Fig-Systematic}{b}, the negative effects induced by the magnetic field fluctuations can be mitigated by an increase in the electric field, which sets the time of the gate implementation. This can be understood by the fact that $\delta B/B_0$ generates an error in the relative phase between the computational states $\ket{0}$ and $\ket{1}$, due to relative errors in $\sigma_{z}$-component of the Hamiltonian as compared to $\sigma_{x}$-component. However, this error becomes smaller when the magnitude of the $\sigma_{x}$-component is increased. 

In the case of fluctuations in the electric field, characterized by a relative error amplitude of $\delta E/E_0$ around the reference value $E_0$, the qubit control is much more robust than in the case of magnetic field fluctuations, as illustrated in Figs.~\SubFig{Fig-Systematic}{c} and~\SubFig{Fig-Systematic}{d}. Nevertheless, our results suggest that choosing appropriately the magnetic field does not allow to reduce the effects of the fluctuating electric field if one is restricted to the range of field amplitude presented in Fig.~\ref{Fig-Anharmonicity}. To understand this result, one needs to keep in mind that $\delta E/E_0$ affects the component $\sigma_{x}$, responsible for changing population of the states $\ket{0}$ and $\ket{1}$, which cannot be corrected by acting only on the relative phase between these states.

\section{Experimental feasibility and challenges}
		
Let us now discuss the feasibility and the challenges to be overcome to implement experimentally the qubit proposed in the present work, and encode information in it. First of all, the fabrication of graphene nanotorus is the main requirement of our proposal of physical qubit, which can then be used to trap the electron. Yet, state-of-art lithography techniques have allowed the creation of nanotori array~\cite{Lehr:14}, which demonstrates the feasibility of designing tori at the proposed scale. As a second step, the trapping of a single electron in the system has to be addressed. Although previous theoretical works suggested that such a task is feasible~\cite{Chan:12}, the experimental realization of atom, ion or electron trapping in graphene surfaces seems to be more challenging than the nanotorus production itself. In fact, beyond the technical issue of this challenge, Klein tunnelling~\cite{Klein:29} appeared as a limiting phenomenon for the trapping performance of relativistic electrons on graphene surfaces~\cite{Gutierrez:16,Renjun:18,Zhang:22}. Finally, the encoding of information requires an excellent control of the electronic states of the system, in order to properly store quantum information. Thus, one needs quantum control at the electronic level. In this context, a recent experiment~\cite{Martinez:22} reported a new strategy based on low-temperature scanning tunneling imaging and spectroscopy to access the molecular structure of graphene nanorings. Although it has been used to characterize the system (such  as the graphene structure), this represents a promising step toward the control of quantum information encoded in the electronic states, including in the present proposal of an electron trapped on the surface of a graphene nanotorus.
		
Once these challenges will be overcome, the encoding of information should become an  efficient process, and other aspects such as the coherence time of the system will have to be investigated. Indeed, an important limitation in the performance of platforms acting as qubits is the decoherence mechanisms at play. For example, differently from the solid-state spin qubits proposed in quantum dots~\cite{Loss:98}, we do not expect the graphene nanotorus to be robust to environmental fluctuations in the electronic structure of the system, since the spin of the system is not used to encode the information. Conversely, the nanotorus may be more robust than spin quantum dot qubits to environmental effects on the system spin, such as transverse and longitudinal spin relaxation~\cite{Sarthour:Book,Khaetskii:20,Woods:02,Zhang:20_spin,lawrie2020spin,Banszerus:22}. At the moment, a proper description the robustness of the graphene nanotorus against decoherence effects remain to be determined, accounting for the different coupling of the system to its surroundings.
	
\section{Conclusions and prospects}
	
In this paper we have introduced a new kind of physical two-level system which can be used as a physical quantum bit. The curvature of a graphene surface induces a confining potential for an electron, which makes the electronic energy level structure suitable to the encoding of quantum information as a superposition of the two lowest energy states $\ket{l=\{0,1\},m=0}=\ket{\{0,1\}}$. Arbitrary single-qubit states of the form $\ket{\psi} = a\ket{0} + b\ket{1}$ can then be prepared and manipulated using external magnetic and electric fields. Our system obeys the DiVincenzo's criteria which state the qubit must allow for a perfect encoding and a coherent control of information. As a first analysis of the effect of errors in the qubit operations, we quantified how undesired fluctuations in the magnetic and electric fields affect the gate fidelity. These errors can be mitigated by increasing the amplitude of the electric field used to implement the single-qubit gates.
	
It is timely to say that any proposal of quantum computer begins with a proposal of highly controllable single qubits. However, the scalability and selective controllability of multi-qubit gates are other fundamental requirements in DiVincenzo's view~\cite{Divincenzo:00}. In particular, a system able to implement arbitrary single-qubit gates and any entangling two-qubit gate (such as the CNOT, or controlled flip phase gate) is eligible as a device able to perform universal quantum computation~\cite{Barenco:95}. In our system, building a two-qubit system is, in principle, possible due to the recent progress in the production and control of multiple organic nanorings~\cite{Chen:05,Thorner:14,Kharissova:19}, along with the progress on their realization on a larger scale~\cite{Lehr:14}. Moreover, understanding the influence of the environment over the system, and thus deccoherence effects~\cite{Petruccione:Book}, is a fundamental issue which remains to be understood. Due to the coupling of electrons with the phonon modes in graphene~\cite{RevModPhys.81.109,PhysRevB.76.045430,PhysRevB.84.035433,park2014electron,PhysRevB.61.10651,Sohier:Thesis}, in addition to the hot-electron relaxation~\cite{Hwang:13} and curvature effects~\cite{PhysRevLett.120.195301,PhysRevB.103.L201104}, the environment is bound to exert an influence over the computational performance of our system. In this sense, our results should not be understood as the whole story, but rather as the initial step of a long and challenging path toward a new platform for quantum computation.
	
	\begin{acknowledgments}
		R.B. is supported by the S\~ao Paulo Research Foundation (FAPESP) through Grants Nos. 2019/22685-1, 2019/13143-0, 2018/15554-5 and 2021/10224-0. R.B. received support from the National Council for Scientific and Technological Development (CNPq) Grant Nos. 313886/2020-2 and 409946/2018-4. A.C.S is supported by the European Union's Horizon 2020 FET-Open project SuperQuLAN (899354), and by the Proyecto Sinérgico CAM 2020 Y2020/TCS-6545 (NanoQuCo-CM) from the Comunidad de Madrid. A.C.S. acknowledges the support by the S\~ao Paulo Research Foundation (FAPESP) through Grants Nos. 2019/22685-1 and 2021/10224-0.
	\end{acknowledgments}

	\appendix
	
	\section{Dimensional analysis of Eq.~\eqref{Eq-Mu}} \label{ApSec-DimenAnaly}
	
	Let us consider
	\begin{align}
	\mu = er \left[\sqrt{\frac{\hbar}{2r|\beta|}} -\frac{1}{6}\sqrt{\left(\frac{\hbar}{2r|\beta|}\right)^3} \right]
	\end{align}
	where the dimensional analysis of $\beta$ follows
	\begin{align}
	[\beta] &= \sqrt{[r]\left[\frac{[\hbar]^2}{[(R-r)]^3}+[e]^2[B]^2[(R-r)]\right]} \nonumber \\
	&= \sqrt{m\left[\frac{(J \cdot s)^2}{m^3}+C^2T^2m\right]} = \sqrt{\frac{J^2 \cdot s^2}{m^2}+C^2 \frac{Kg^2}{C^2 \cdot s^2} m^2} \nonumber \\
	&= \sqrt{\frac{J^2 \cdot s^2}{m^2}+\frac{Kg^2}{s^2} m^2}= \sqrt{\frac{Kg^2 \cdot m^4}{s^4}\frac{s^2}{m^2}+\frac{Kg^2}{s^2} m^2} \nonumber \\
	&= \sqrt{\frac{Kg^2 \cdot m^2}{s^2}} = \frac{Kg \cdot m}{s}
	\end{align}
	such that
	\begin{align}
	\frac{[\hbar]}{[r][|\beta|]} &= \frac{J \cdot s}{m}\frac{s}{Kg \cdot m} =J\frac{s^2}{Kg \cdot m^2} \nonumber \\
	&= \frac{Kg \cdot m^2}{s^2}\frac{s^2}{Kg \cdot m^2} = 1 ,
	\end{align}
	therefore we get
	\begin{align}
	[\mu_{0}] = [e][r] = C\cdot m ,
	\end{align}
	the dimension of electric dipole moment.
	
	\section{Effective dynamics} \label{ApSec-EffecDyn}
	
	Consider the total Hamiltonian $H(t)=H_{0} + H_{1}(t)$, in which $H_{0} =\hbar \omega \sigma_{+}\sigma_{-}$, as
	\begin{align}
	H(t) = \hbar \omega \sigma_{+}\sigma_{-} + \hbar \Omega(E_0) \cos( \revisionRefB{\omega_{\mathrm{d}}} t + \phi) \left( \sigma_{-} + \sigma_{+} \right)  .
	\end{align}
	
	Since in the above equation we have an small highly oscillating term, it is convenient to identify how to approximate it to an effective Hamiltonian using the Rotating Wave Approximation (RWA). To this end, we write the Schrödinger equation in the interaction picture using the transformation $\ket{\psi_{\text{I}}(t)} = e^{iH_{0}t/\hbar} \ket{\psi(t)}$ we get the dynamics for $\ket{\psi_{\text{I}}(t)}$ as 
	\begin{align}
	\ket*{\dot{\psi}_{\text{I}}(t)} = i\hbar V_{\text{I}}(t) \ket{\psi_{\text{I}}(t)} ,
	\end{align}
	with $V_{\text{I}}(t) = e^{iH_{0}t/\hbar} H_{1}(t) e^{-iH_{0}t/\hbar}$. As a first remark, note that $[H_{1}(t_{1}),H_{1}(t_{2})]=[V_{\text{I}}(t_{1}),V_{\text{I}}(t_{2})]=0$ for any $(t_1,t_2)$. It allows to write the solution for the system dynamics as
	\begin{align}
	\ket{\psi_{\text{I}}(t)} = U_{\text{I}}(t) \ket{\psi_{\text{I}}(0)} ,
	\end{align}
	with
	\begin{align}
	U_{\text{I}}(t) = \exp \left[ \frac{1}{i\hbar} \int_{0}^{t} V_{\text{I}}(\xi) d\xi \right] . \label{AppUI}
	\end{align}
	
	We use that
	\begin{align}
	V_{\text{I}}(t) &= e^{iH_{0}t/\hbar} H_{1}(t) e^{-iH_{0}t/\hbar} \nonumber \\
	&= \frac{\hbar \Omega(E_0)}{2} \left( e^{i(\revisionRefB{\omega_{\mathrm{d}}} t + \phi)} + e^{-i(\revisionRefB{\omega_{\mathrm{d}}} t + \phi)}\right) e^{iH_{0}t/\hbar}(\sigma_{+} + \sigma_{-})e^{-iH_{0}t/\hbar} \nonumber \\
	&= \frac{\hbar \Omega(E_0)}{2} \left[\left( e^{i(\omega_{-} t + \phi)} + e^{-i(\omega_{+} t + \phi)}\right) \sigma_{-} + \mathrm{H.c.}\right] ,
	\end{align}
	where $\omega_{\pm} = \revisionRefB{\omega_{\mathrm{d}}} \pm \omega$ and `H.c.' denotes the Hermitian conjugate. Therefore, by integrating the above expression
	\begin{align}
	\int_{0}^{t} V_{\text{I}}(\xi) d\xi &= \frac{i\hbar\Omega(E_0)}{2} \left[ \frac{e^{-i \phi} \left( e^{-i\omega_{+} t} -1 \right)}{\omega_{+}} - \frac{e^{i \phi} \left( e^{i\omega_{-} t} -1 \right)}{\omega_{-}} \right] 
	\sigma_{-} \nonumber \\
	&+\frac{i\hbar\Omega(E_0)}{2}\left[\frac{e^{-i \phi} \left( e^{-i\omega_{-} t} -1 \right)}{\omega_{-}} -\frac{e^{i \phi} \left( e^{i\omega_{+} t} -1 \right)}{\omega_{+}} \right] 
	\sigma_{+} .
	\end{align}
	
	Thus, for a highly oscillating external field, we can approximate the above expression to
	\begin{align}
	\int_{0}^{t} V_{\text{I}}(\xi) d\xi \approx \frac{\hbar\Omega(E_0)}{2} \left[ 
	\frac{-ie^{i \phi} \left( e^{i\omega_{-} t} -1 \right)}{\omega_{-}} \sigma_{-} + \mathrm{H.c.} \right] .
	\end{align}
	
	For this reason, in our analysis we can neglect highly oscillating terms. In this case, we follow by writing the Hamiltonian $H(t)$ in the rotating frame by using $R(t) = e^{i\revisionRefB{\omega_{\mathrm{d}}} t\sigma_{+}\sigma_{-}}$, such that we find
	\begin{align}
	H_{R} &= R(t) H(t) R^{\dagger}(t) + i\hbar \dot{R}(t)R^{\dagger}(t) = \hbar \Delta \sigma_{+}\sigma_{-} + R(t) H_{1}(t) R^{\dagger}(t) \nonumber \\ 
	&= \hbar \Delta \sigma_{+}\sigma_{-} + \frac{\hbar \Omega(E_0)}{2} \left( e^{i(\revisionRefB{\omega_{\mathrm{d}}} t + \phi)} + e^{-i(\revisionRefB{\omega_{\mathrm{d}}} t + \phi)}\right) e^{-i \revisionRefB{\omega_{\mathrm{d}}} t} \sigma_{-} + \mathrm{H.c.} ,
	\end{align}
	where we define $\Delta = \omega - \revisionRefB{\omega_{\mathrm{d}}}$. Now, by neglecting the highly oscillating terms $e^{\pm 2i\revisionRefB{\omega_{\mathrm{d}}} t}$, it is possible to proof that we get
	\begin{align}
	H_{R} = \hbar \Delta \sigma_{+}\sigma_{-} + \frac{\hbar \Omega(E_0)}{2} \left( e^{i\phi} \sigma_{-} + e^{-i\phi} \sigma_{+}\right) .
	\end{align}

\end{document}